\documentclass[twocolumn, switch]{article} % Method A for two-column formatting

\usepackage{preprint}
\usepackage{algorithm}
\usepackage{algorithmic}
\usepackage{amsmath, amsthm, amssymb, amsfonts}
\usepackage{graphicx}
\usepackage{xcolor}
\usepackage{amsmath}
\usepackage{enumerate}
\usepackage{booktabs}
\usepackage{multirow}
\usepackage{tabularx}
\usepackage{subcaption}

\usepackage[numbers,square]{natbib}
\usepackage[utf8]{inputenc}	% allow utf-8 input
\usepackage[T1]{fontenc}	% use 8-bit T1 fonts
\usepackage{xcolor}		% colors for hyperlinks
\usepackage[colorlinks = true,
            linkcolor = purple,
            urlcolor  = blue,
            citecolor = cyan,
            anchorcolor = black]{hyperref}	% Color links to references, figures, etc.
\usepackage{booktabs} 		% professional-quality tables
\usepackage{nicefrac}		% compact symbols for 1/2, etc.
\usepackage{microtype}		% microtypography
\usepackage{lineno}		% Line numbers
\usepackage{float}			% Allows for figures within multicol

\usepackage{lipsum}		%  Filler text

\usepackage{newfloat}
\DeclareFloatingEnvironment[name={Supplementary Figure}]{suppfigure}
\usepackage{sidecap}
\sidecaptionvpos{figure}{c}

\usepackage{titlesec}
\titlespacing\section{0pt}{12pt plus 3pt minus 3pt}{1pt plus 1pt minus 1pt}
\titlespacing\subsection{0pt}{10pt plus 3pt minus 3pt}{1pt plus 1pt minus 1pt}
\titlespacing\subsubsection{0pt}{8pt plus 3pt minus 3pt}{1pt plus 1pt minus 1pt}

\title{Field Demonstration of a Multi-User Continuous-Variable Quantum Access Network for Quantum-to-the-Home}

\usepackage{eso-pic}
\usepackage{tikz}
\usepackage{xcolor}
\PassOptionsToPackage{colorlinks=true,linkcolor=gray,urlcolor=gray}{hyperref}

\newcommand{\AddMyWatermarks}{%
  \begin{tikzpicture}[remember picture, overlay]
    \node[rotate=90, color=gray!60, scale=1] at ([xshift=-4.05in,yshift=0in]current page.center) {%
      \href{https://doi.org/}{Publication doi}%
    };
    \node[rotate=90, color=gray!60, scale=1] at ([xshift=3.9in,yshift=0in]current page.center) {%
      \href{https://doi.org/}{Preprint doi}%
    };
    \node[color=gray!90, scale=1] at ([xshift=0in,yshift=-5in]current page.center) {%
      This is the author's accepted manuscript. The final version will appear in XXXX 2025, © XXXX 2025.%
    };
  \end{tikzpicture}%
}

\AddToShipoutPictureBG*{\AddMyWatermarks}

\usepackage{titling}
\usepackage{orcidlink}
\usepackage{footmisc}
\newcommand{\Author}[3]{% Name, ORCID, Institution
  \textbf{#1}\textsuperscript{#2}%
}

\author{
  \Author{Junpeng Zhang}{1,$\dagger$}{0000-0000-0000-0000} \and
  \Author{Xu Liu}{1,$\dagger$}{0000-0000-0000-0000}\and
  \Author{Qijun Zhang}{1,$\dagger$}{0000-0000-0000-0000}\and
  \Author{Yifeng Liang}{1}{0000-0000-0000-0000} \and
  \Author{Yue Yu}{1}{0000-0000-0000-0000} \and
  \Author{Peng Huang}{1,2,3}{0000-0000-0000-0000}\and
  \Author{Huasheng Li}{4}{0000-0000-0000-0000} \and
  \Author{Yingming Zhou}{4}{0000-0000-0000-0000}\and \Author{Jingyu Yang}{5}{0000-0000-0000-0000} \and
  \Author{Chunchen Li}{5}{0000-0000-0000-0000} \and
  \Author{Yunfan Chen}{5}{0000-0000-0000-0000}\and
  \Author{Cheng Zheng}{5}{0000-0000-0000-0000} \and
  \Author{Ciqing Deng}{5}{0000-0000-0000-0000} \and
  \Author{Tao Wang}{1,2,3}{0000-0000-0000-0000} \and
  \Author{Guihua Zeng}{1,2,3,4}{0000-0000-0000-0000}
}

\date{%
  \textsuperscript{1}State Key Laboratory of Photonics and Communications, Center of Quantum Sensing and Information Processing, Shanghai Jiao Tong University, Shanghai 200240, China\\
  \textsuperscript{2}Shanghai Research Center for Quantum Sciences, Shanghai 201315, China\\
  \textsuperscript{3}Hefei National Laboratory, Hefei, 230088, China\\
  \textsuperscript{4}Shanghai XunTai Quantech Co., Ltd, Shanghai, 200241, China\\
  \textsuperscript{5}China Telecom Corporation Limited Shanghai Branch, Shanghai, 200041, China\\
  \textsuperscript{$\dagger$}The authors contributed equally to this work\\[1em]
  \footnotesize \textbf{Corresponding author:} Tao Wang\texttt{<tonystar@sjtu.edu.cn>}\\
}

\begin{document}

\twocolumn[ % Method A for two-column formatting
  \begin{@twocolumnfalse} % Method A for two-column formatting

\maketitle
\thispagestyle{empty}

\begin{abstract}
	Realizing scalable Quantum-to-the-Home (QTTH) faces a bottleneck: link asymmetry in broadcast continuous-variable quantum access networks (CV-QANs) hinders the selection of a globally optimal modulation variance. We demonstrate a downstream broadcast CV-QAN connecting a Quantum Line Terminal (QLT) to multiple Quantum Network Units (QNUs) over commercial fiber. Operating within a trusted local network domain, we establish a multi-user utility model to select the optimal shared variance, balancing network efficiency and user fairness. Supported by robust digital signal processing, our 1:16 field trial achieves Mbit/s-level asymptotic secure key rates, bridging theoretical protocols with Fiber-to-the-Home reality and guiding future scalable access architectures.
\end{abstract}
%\keywords{First keyword \and Second keyword \and More} % (optional)
\vspace{0.35cm}

  \end{@twocolumnfalse} % Method A for two-column formatting
] % Method A for two-column formatting

%\begin{multicols}{2} % Method B for two-column formatting (doesn't play well with line numbers), comment out if using method A

%%%%%%%%%%%%%%%  Main text   %%%%%%%%%%%%%%%
% \linenumbers

\section{Introduction}
Quantum key distribution (QKD) is a communication technology that exploits the principles of the no-cloning theorem and measurement-induced disturbance of quantum states to securely distribute keys\cite{wehner2018quantum,pirandola2020advances,gisin2002quantum,buchmann2017quantum,bennett2014quantum}. By providing a theoretically unconditionally secure key-sharing mechanism at the physical layer for communicating parties, it guarantees information-theoretic security for network communications, making it one of the most successful applications of quantum cryptography. QKD networks represent a critical stage in the development of the quantum internet\cite{wootters1982single,hall1995information,kimble2008quantum}, breaking through the physical constraints of single links to achieve quantum-secure access for massive terminals and the sharing of key resources. Evolving from early point-to-point QKD systems, QKD networks enable key sharing among multiple users through the preparation and measurement of quantum states. Depending on the physical encoding of information, QKD protocols can be broadly classified into discrete-variable (DV-QKD)\cite{bennett2014quantum,lo2005decoy,lo2012measurement,lucamarini2018overcoming,zhao2006experimental,boaron2018secure,zhang2022device,koashi2009simple,scarani2009security} and continuous-variable (CV-QKD)\cite{polkinghorne1999continuous,grosshans2002continuous,jouguet2013experimental,ralph2000security,grosshans2002reverse,grosshans2003quantum,weedbrook2012gaussian,zhang2020long,leverrier2015composable,tian2022experimental} schemes. The DV scheme dominates long-haul backbone and metropolitan networks by leveraging single-photon detection , while the CV scheme utilizes coherent detection to exhibit exceptional compatibility with classical wavelength-division multiplexing systems , making it a compelling implementation scheme for quantum access networks in recent years.

Driven by the continuous maturity of QKD protocols and hardware ecosystems, QKD technology has transitioned from the laboratory to large-scale practical infrastructure construction \cite{chen2010metropolitan,wang2010field,zhang2018large,tang2016measurement,bacco2019field,jouguet2012field}.Consequently, quantum backbone and metropolitan area networks are being progressively deployed worldwide , exemplified by the Beijing-Shanghai backbone\cite{chen2021implementation},and various field-deployed metropolitan networks in Shanghai\cite{huang2016field}, Geneva\cite{Stucki2011long}, Tokyo\cite{sasaki2011field},Cambridge\cite{dynes2019cambridge}, and other global locations.

\begin{figure*}[htbp]
	\centering
	\includegraphics[width=1.0\textwidth]{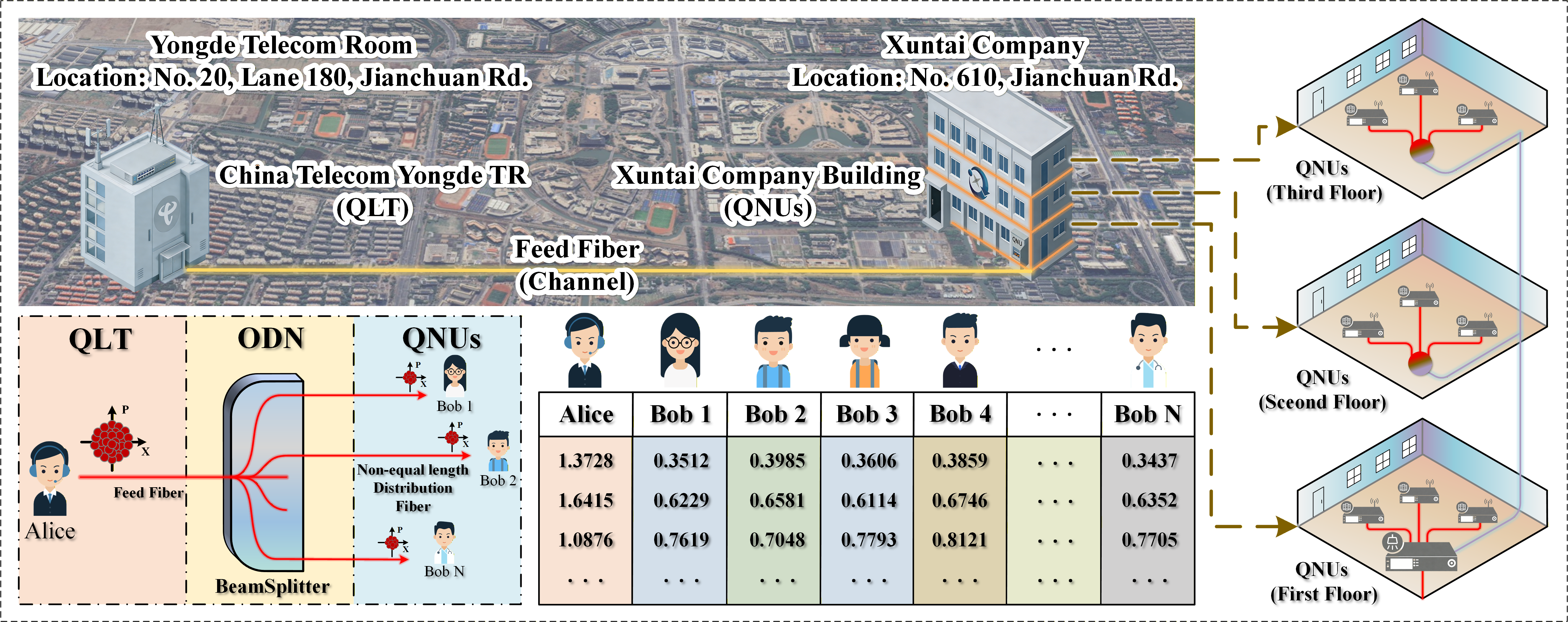} % width通常设为0.8到1.0倍页面宽度
	\caption{Field-deployed downstream broadcast CV-QANs and its practical multi-user model. The upper panel shows the real deployment scenario between the China Telecom Yongde telecom room (QLT) and the XunTai company building, connected by a feeder fiber. The right panel illustrates the floor-wise distribution of quantum network units (QNUs) inside the building, where unequal indoor drop-fiber lengths result in asymmetric channel conditions across users. The lower-left panel presents the corresponding downstream broadcast architecture, in which Alice transmits Gaussian-modulated quantum signals through the ODN to multiple Bobs. The lower-middle panel shows representative correlated quadrature data shared by Alice and different users. This figure highlights the practical access-network scenario considered in this work, where a common transmitter-side configuration is imposed on all users despite unequal channel attenuation and unbalanced SKR.}
	\label{field test}
\end{figure*}

However, distinct from backbone and metropolitan infrastructure, quantum access networks (QANs) directly interface with diverse end-user premises (such as homes, buildings, campuses, and enterprises), bridging the gap between large-scale deployment and widespread practical application. By integrating QKD into P2MP passive optical networks (PONs), QANs serve as this critical "last-mile" link\cite{altabas2018passive}. Consequently, downstream multi-user QKD and its coexistence with classical Fiber-to-the-Home (FTTH) services have been extensively investigated\cite{townsend1997quantum,choi2011quantum,wang2021practical,huang2024cost,liu2024integrated,xu2023round,wang2023experimental}. Among these, CV-QANs utilizing a broadcast scheme have emerged as highly representative. Recent studies have theoretically proposed and experimentally verified P2MP QKD protocols based on coherent state broadcast \cite{bian2023high,hajomer2024continuous}, including a 16-node high-rate QAN laboratory demonstration achieving Mbit/s-level secret key rates per user\cite{pan2025high}.

In this paper, we demonstrate a CV-QAN in a commercial telecommunication fiber environment, providing practical evidence for Quantum-to-the-Home (QTTH) by demonstrating that quantum-secure communication can be extended to "last-mile" access scenarios based on traditional FTTH architectures. As illustrated in Fig.
\ref{field test}, the system is deployed on China Telecom's optical fiber infrastructure in the Minhang District of Shanghai, following a tree-topology access architecture. The quantum line terminal (QLT) located at the Yongde telecommunication room (TR) distributes quantum signals over an approximately 4-km feeder fiber to the XunTai building, where an optical distribution network (ODN) fans out the signals to 16 quantum network units (QNUs), forming an operational multi-user quantum access network.

Beyond field deployment, this work addresses inherent challenges of downstream broadcast architectures. Realistic split fibers introduce significant asymmetric channel , yielding imbalanced secret key rates typically bottlenecked by the weakest link. Furthermore, because modulation precedes distribution, the modulation variance becomes a system-wide shared parameter, precluding individual optimization and creating tension across asymmetric channels.

To address this issue, we establish a system model and an evaluation methodology that account for asymmetric, realistic optical fiber channels. Based on the constructed mathematical model, we propose an optimal selection strategy for the modulation variance in multi-user systems. These results provide deployment-relevant guidance and performance bounds for CV-QANs.t-relevant guidance and performance bounds for CV-QANs.

\section{Scheme}
\label{sec:Scheme} 

\subsection{CV-QKD and security key rate}
In a standard P2P Gaussian-modulated coherent state (GMCS) CV-QKD system, information is encoded onto the optical field quadratures. Under the framework of asymptotic security with reverse reconciliation, the secret key rate (SKR) is conventionally quantified as\cite{weedbrook2012gaussian}:

\begin{equation}
	\mathrm{SKR}_{AB} = f \left( \beta I(A:B) - \chi_{BE} \right)
\end{equation}

where $f$ is the system repetition frequency, and $\beta \in [0,1]$ is the reconciliation efficiency. The term $I(A:B)$ denotes the classical mutual information between Alice and Bob, which is governed by the channel transmittance $T$, excess noise $\xi$, and modulation variance $V_A$. Finally, $\chi_{BE}$ represents the Holevo bound defining the maximum information accessible to an eavesdropper (Eve).
efining the maximum information accessible to an eavesdropper (Eve).

\subsection{Broadcast based CV-QPONs}
\label{subsec:Broadcast} 
Evolving from P2P to P2MP architectures, downstream CV-QPONs frequently employ the broadcast quantum state scheme by capitalizing on the passive splitting characteristics of ODN. Specifically, a single GMCS signal generated by Alice is physically divided by a $1:N$ ODN and simultaneously distributed to all $N$ receivers ($\text{Bob}_1, \dots, \text{Bob}_i, \dots, \text{Bob}_N$).

This physical power splitting inherently introduces user crosstalk among receiving nodes, necessitating specific security modeling. Recent studies have handled this crosstalk based on different assumptions regarding user behavior: either by accounting for potential collaboration where Eve exploits other users' leaked information\cite{bian2026approaching}, or by assuming strict non-collaboration where participants act entirely independently of Eve \cite{pan2025high}. In this work, we consider the strict non-collaboration setting, which is well-suited for our field-deployed access scenario operating within a single administrative domain. Specifically, all 16 QNUs are deployed within the same building environment, and the corresponding security analysis is therefore carried out under the assumption that the coexisting users do not collaborate with Eve. Under this assumption, Pan et al.\cite{pan2025high} established the corresponding security analysis, and the broadcast key rate can be written as:

\begin{equation}
	\mathrm{SKR}_{AB_i}^\text{Broadcast} = f \biggl( \beta I(A;B_{i}) 
	- \max I(B_{i};\bar{\mathbf{B}}_{N}^{i}),\chi_{B_{i}E}^{(N)}(\gamma_{A\mathbf{B}_{N}})
	\biggr)
\end{equation}

To satisfy both the security requirement against Eve and the key independence requirement among users, it suffices to remove the maximum value between the two. Under typical experimental conditions, the contribution of $\chi_{B_{i}E}^{(N)}(\gamma_{A\mathbf{B}_{N}})$ is generally greater than that of $I(B_{i};\bar{\mathbf{B}}_{N}^{i})$ , so the key rate formula simplifies to the conventional expression:

\begin{equation}
	\mathrm{SKR}_{AB_i}^\text{Broadcast}=f \left(\beta I(A;B_{i})-\chi_{B_{i}E}^{(N)}(\gamma_{A\mathbf{B}_{N}})\right)
	\label{eq:SKR}
\end{equation}
Where $I(A;B_{i})$ denotes the classical mutual information between Alice and $\text{Bob}_i$, while the term $\chi_{B_{i}E}^{(N)}(\gamma_{A\mathbf{B}_{N}})$ signifies the Holevo information potentially accessible to Eve.

\subsection{Optimal modulation variance selection under practical channel conditions}
The fundamental distinction between field-deployed quantum access networks and laboratory-based experimental setups lies in the asymmetry of the optical channels experienced by different users.
In practical field implementations, each user is connected through an independently deployed fiber infrastructure, such as FTTH, which inevitably leads to different channel lengths and losses for different Bobs with respect to the transmitter Alice.

In downlink broadcast quantum access networks, the modulation variance is typically constrained to be a system-level parameter shared by all users, rather than being individually optimized for each Bob. Under this practical constraint, although each user would in principle possess an individual optimal modulation variance determined by its channel transmittance, the strong dependence of the SKR on the channel conditions renders the globally optimal choice of the modulation variance inherently influenced by the user-specific channel asymmetry.

This channel-length-induced asymmetry motivates the investigation of modulation variance optimization strategies in multi-user QANs, which will be discussed in the following subsections.

\subsubsection{Unified system-utility framework for asymmetric channels}
To characterize SKR optimization in CV-QANs under asymmetric channel conditions from a system-level perspective, we formulate the selection of the shared modulation variance as a network-wide utility maximization problem. On this basis, a unified $\alpha$-fair system utility function, $\text{U}_{sys}$ , is introduced to quantify how well a global modulation variance $V_A$ matches the asymmetric channel conditions experienced by different users. The utility is defined as

\begin{equation}
	\text{U}_{sys}(V_A, \alpha) =
	\begin{cases}
		\displaystyle \sum_{j=1}^{N} \frac{\text{SKR}_j(V_A)^{1-\alpha}}{1-\alpha}, & \alpha \ge 0,\alpha \neq 1 \\[12pt]
		\displaystyle \sum_{j=1}^{N} \ln(\text{SKR}_j(V_A)), & \alpha = 1
	\end{cases}
\end{equation}

where $\text{SKR}_j(V_A)$  denotes the asymptotic SKR of the $j$-th user under the shared modulation variance  $V_A$ . The parameter $\alpha \ge 0$  characterizes the degree of sensitivity to low-rate users during modulation-variance selection, and can also be interpreted as the extent to which the system emphasizes the performance of disadvantaged users. Accordingly, the optimal shared modulation variance over the physically meaningful range $\mathcal{V}$ can be written in a unified form as
\begin{equation}
	V_A^*(\alpha) = \arg\max_{V_A \in \mathcal{V}} \text{U}(V_A, \alpha),
\end{equation}
Here, the operator $\arg\max$ denotes the specific value of $V_A$ within the feasible set $\mathcal{V}$ that maximizes the utility function $\text{U}(V_A, \alpha)$.

In this framework, the modulation variance is treated as the key shared parameter governing the overall secret-key-rate distribution across the network, while different values of $\alpha$ correspond to different operational preferences. When $\alpha$ is small, $\text{U}_{sys}$ places greater emphasis on aggregate SKR, corresponding to an efficiency-oriented operating regime. As $\alpha$ increases, the utility function becomes progressively more sensitive to low-rate users, and the optimization correspondingly shifts toward protecting weak-user performance, thereby enhancing fairness. At $\alpha=1$ the utility reduces to a logarithmic summation form, which corresponds to proportional fairness and provides a balanced trade-off between overall efficiency and user equity. Therefore, the optimization of the shared modulation variance can be interpreted in a unified manner as selecting an appropriate CV-QAN operating point along the efficiency–fairness trade-off in a practical asymmetric access-network environment.

\begin{figure}[htbp]
	\centering
	\includegraphics[width=\linewidth]{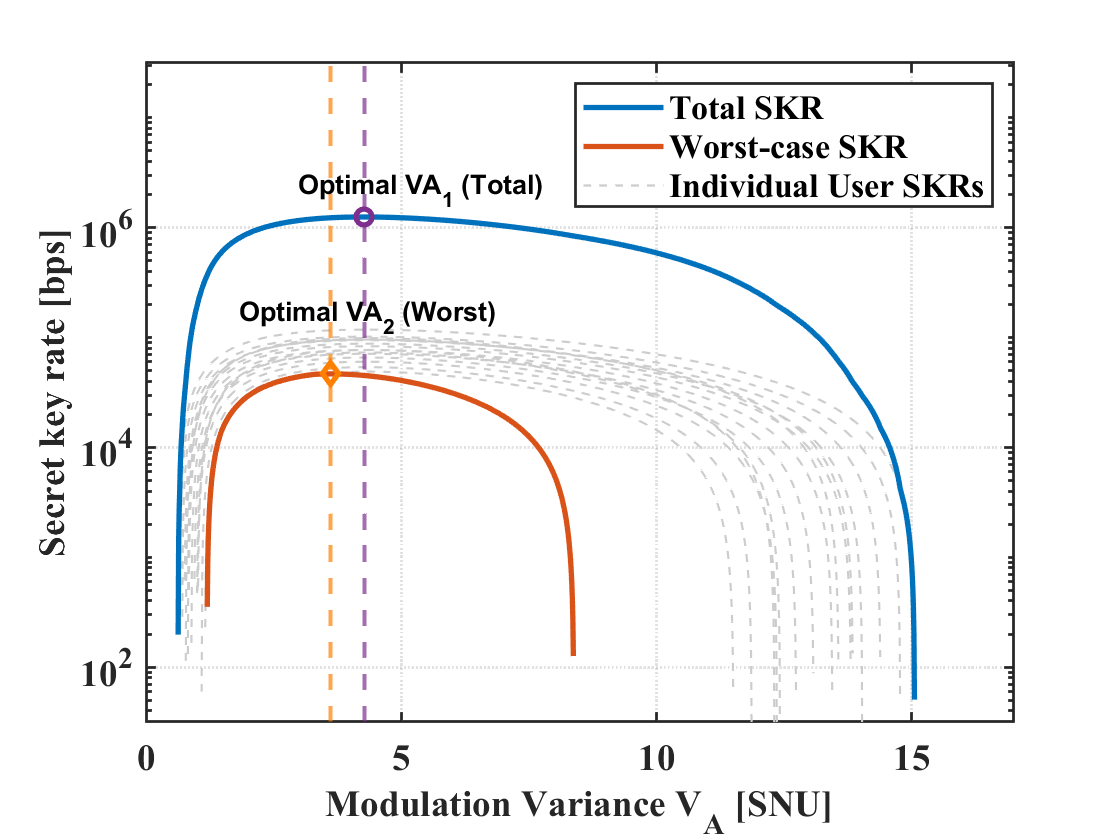} 
	\caption{SKR versus shared modulation variance under asymmetric access-channel conditions. The blue curve shows the sum SKR of all users versus the shared modulation variance $V_A$ , corresponding to sum-SKR maximization in the $\alpha \to 0$ limit, while the red curve shows the minimum user SKR, corresponding to max–min optimization in the $\alpha \to \infty$ limit. Gray dashed curves denote the SKRs of individual users. The two objectives yield different optima,  indicating that a single shared$V_A$cannot simultaneously optimize both aggregate key supply and worst-user serviceability.}
	\label{fangcha}
\end{figure}

\subsubsection{Sum-SKR Maximization in the $\alpha \to 0$ limit}
In the limit of $\alpha \rightarrow 0$, the proposed system utility $\text{U}_{sys}$ reduces to the linear summation of the SKRs of all users. Under this limit, the secure keys generated across the network are treated as directly additive key resources, and the objective of selecting the optimal modulation variance becomes the maximization of the aggregate key-generation capability of the access network rather than the equalization of user-level performance. 

\begin{equation}
	\text{U}_{sys}(V_A, \alpha) = \sum_{j=1}^{N} \text{SKR}_j(V_A), \alpha = 0
\end{equation}

This limiting case corresponds to an efficiency-oriented operating regime in which the system primarily seeks to maximize the total amount of extractable secure key material per unit time. In this regime, no additional priority is assigned to the protection of low-rate users.

Consequently, the optimal shared modulation variance is typically determined to a greater extent by the contribution of high-quality links, reflecting a resource-allocation strategy centered on maximizing the overall key-supply capability of the network. As illustrated by the blue curve in Fig. \ref{fangcha}, the aggregate SKR reaches its maximum at optimal $V_{A1}$. From a physical perspective, this efficiency-oriented optimum corresponds to an operating point dominated by high-rate links. Because such links can continue to provide substantial key contributions under relatively strong modulation, the maximizer of the sum SKR is generally shifted toward a larger modulation variance. Put differently, when system efficiency is prioritized, the network may tolerate pronounced performance disparity among users, provided that the aggregate key output continues to improve.

\subsubsection{Min-SKR Maximization in the $\alpha \to \infty$ Limit}
\label{subsubsec:Min-SKR} 
Conversely, as $\alpha$ increases, the proposed utility $\text{U}_{sys}$ places progressively greater emphasis on low-rate users, and the selection of the shared modulation variance becomes increasingly constrained by the security margin of the weakest channels. In the limit of $\alpha \rightarrow \infty$, the optimization reduces to a max-min criterion, where the objective of selecting the optimal modulation variance becomes the maximization of the minimum user SKR:

\begin{equation}
	V_A^*(\alpha) = \arg\max_{V_A }\min_{j} \text{SKR}_j(V_A),
\end{equation}

In this regime, system operation is effectively governed by the worst link, corresponding to a highly conservative resource-allocation strategy. The primary concern is ensuring that the weakest user remains within a secure operating region, while aggregate throughput becomes a secondary consideration. This limit is therefore particularly relevant to scenarios in which service coverage, user reachability, and system robustness are prioritized.

As depicted by the red curve in Fig. \ref{fangcha}, the minimum-user SKR exhibits a clear rise-and-fall trend, reaching its maximum at optimal $V_{A2}$. Compared to the efficiency-oriented operating point ($V_{A1}$), this optimum is shifted noticeably toward a smaller modulation variance. From a physical perspective, weak-link users generally experience lower channel transmittance. Once the modulation variance becomes excessively large, the corresponding increase in Eve's accessible information ($\chi_{B_iE}$) degrades the SKR of low-transmittance users earlier than that of stronger links. Consequently, although stronger modulation may still enhance the key contribution of high-quality channels, the weakest user's SKR has already begun to decline, thereby lowering the minimum service guarantee of the system. Under the max-min SKR criterion, the optimal modulation variance is therefore fundamentally constrained to a smaller value.

This divergence in optimal operating points indicates that, under asymmetric access-channel conditions, there is generally no single shared modulation variance $V_A$ that is simultaneously optimal for all system objectives. A larger modulation variance is more effective in unlocking the SKR potential of high-quality links and thus increasing aggregate network throughput, whereas a smaller modulation variance is more beneficial for sustaining the SKR of weak-link users and improving the performance of the bottleneck user.Ultimately, the configuration of $V_A$ necessitates a deliberate trade-off between system efficiency and user fairness.

To ensure a rigorous minimum service guarantee across the deployed network, our subsequent experimental demonstrations and performance evaluations will specifically adopt this max-min ($\alpha \rightarrow \infty$) limit as the operational baseline.

\section{Experimental investigation}

\subsection{Experimental setup}
\begin{figure*}[htbp]
	\centering
	\includegraphics[width=0.9\textwidth]{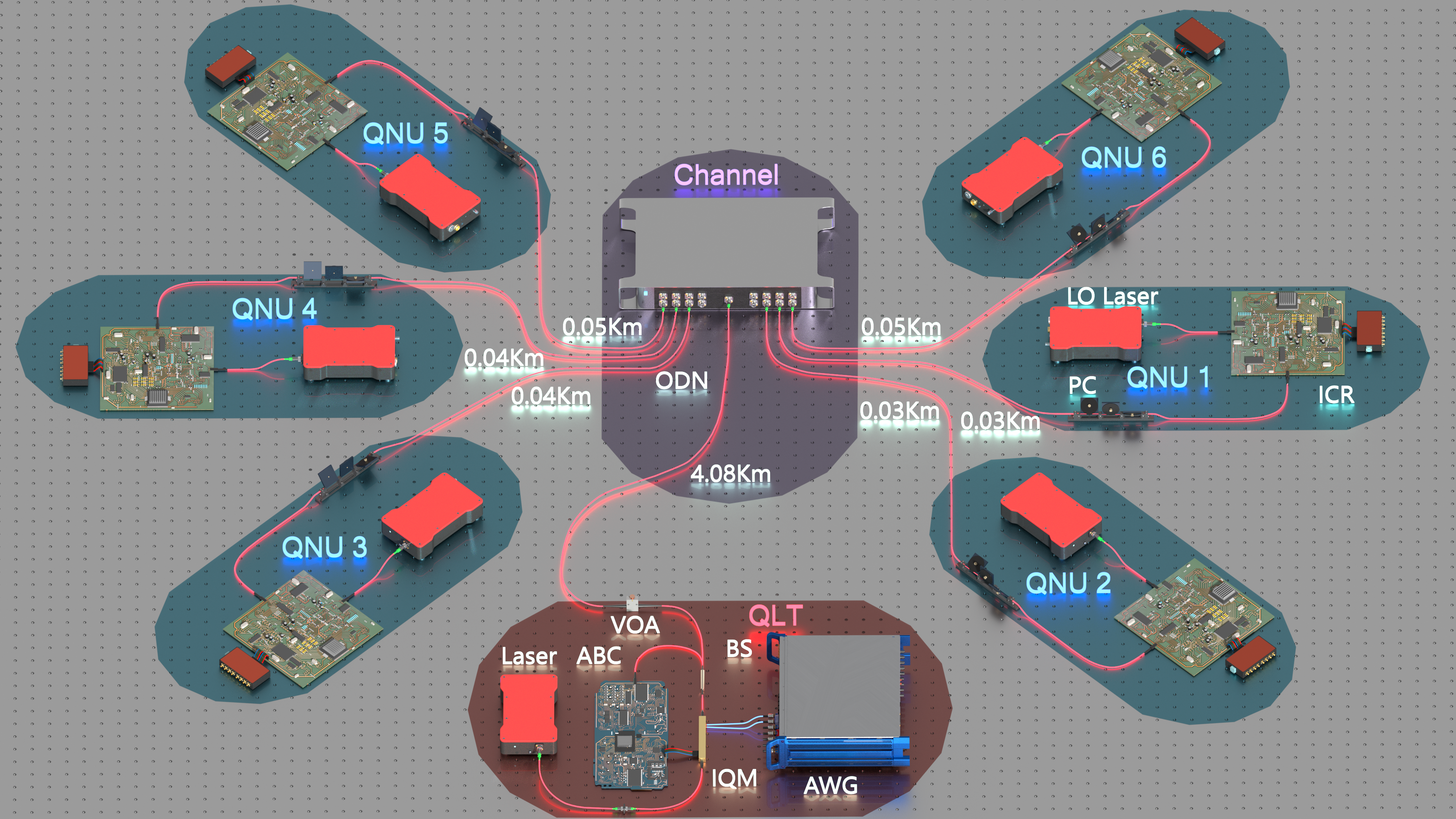} % width通常设为0.8到1.0倍页面宽度
	\caption{Optical setup of the downstream broadcast CV-QANs. The QLT generates GMCS and sends them through a 4.08km feeder fiber to the ODN, which passively distributes the signals to multiple QNUs. Different indoor drop-fiber lengths are labeled for different users. Each QNU performs coherent detection using an LO laser, PC, and ICR. The figure depicts the optical-path implementation of the field experiment.}
	\label{setup}
\end{figure*}

Fig. \ref{setup} illustrates the experimental optical-path schematic of the field-deployed 1-to-16 access network, which comprises a QLT serving as Alice, a PON, and 16 QNUs serving as Bobs. For the security analysis reported in this work, we focus on six representative QNUs that capture the asymmetric channel conditions among the deployed users.

\subsubsection{Quantum Line Terminal (Alice)}
At the QLT, a narrow-linewidth (~100 Hz) laser at $\lambda_A$ = 1550 nm serves as the optical source. The light passes through an isolator into a commercial IQ modulator driven by a 16-bit arbitrary waveform generator (AWG) at a 5 GHz sampling rate. To generate Gaussian-modulated root-raised-cosine (RRC) signals, a 500 MHz random sequence is upsampled to 5 GHz and shaped by an RRC filter (roll-off 0.2), yielding a 600 MHz signal bandwidth. This signal is frequency-shifted to a carrier of $f_{\text{carrier}}$ = -500 MHz, multiplexed with a pilot tone at $f_{\text{pilot}}$ = -1.25 GHz, and attenuated by a variable optical attenuator (VOA). 

For hardware synchronization, a 1490 nm laser generates a 10 MHz amplitude-modulated clock signal, which is wavelength-division multiplexed (WDM) with the quantum signal. Both are launched into the PON, which consists of a 4.08 km single-mode fiber (SMF), a 1$\times$16 beam splitter, and 0.03–0.05 km distribution fibers.

\subsubsection{Quantum Network Unit (Bob)}
The QNUs employ a local local oscillator (LLO) scheme for heterodyne detection. The received signal passes through a polarization controller (PC) into an integrated coherent receiver (ICR) to interfere with the LLO. The orthogonal electrical outputs are digitized at 5 GHz. Concurrently, a photodetector demultiplexes the 10 MHz clock signal, ensuring strict synchronization between the receiver's oscilloscope and the QLT's AWG. Detection data are processed in frames of $10^7$ pulses. Quantum symbols extracted via digital signal processing (DSP) then undergo reverse reconciliation and privacy amplification to yield the final key.

\subsubsection{Digital signal processing}
The DSP algorithm primarily comprises the following steps. 

(1) \textbf{Frequency offset recovery:} In LLO CV-QKD systems, the transmitter and receiver employ independent, free-running lasers. Because the central frequencies of these independent lasers cannot be perfectly identical and will randomly drift with environmental variations, substantial time-varying frequency offsets inevitably exist in the received signal. To address this issue, the system typically multiplexes a high-intensity pilot tone at the transmitter. By precisely searching for and calibrating the spectral peak position of the pilot tone in the frequency domain, the real-time frequency offset $\Delta f$ can be calculated. Subsequently, the received signal undergoes corresponding spectrum shifting to perform frequency offset compensation. This effectively eliminates the frequency drift of the signal, ensuring its spectral center precisely returns to the predetermined frequency.

(2) \textbf{Phase recovery:} In addition to the fixed frequency difference, disturbances within the practical optical fiber channel also induce phase drift. Such phase drift causes the quantum signal to randomly rotate in the phase space, which directly increases the excess noise of the system and degrades the SKR. The core of phase recovery lies in utilizing the pilot tone to establish a stable phase reference. Because the quantum signal and the pilot tone co-propagate through the same optical fiber link and experience identical phase drift, the extracted phase information can be directly applied to the quantum signal, thereby achieving precise calibration and compensation of the quantum state's phase

(3) \textbf{Coherent demodulation} Upon completing comprehensive frequency and phase synchronization, the signal must be demodulated to extract the final measurement data. First, digital spectral shifting techniques are employed to down-convert the quantum signal from a specific frequency band to the baseband. This is immediately followed by a digital low-pass filtering operation to thoroughly eliminate out-of-band noise and residual interference from the pilot tone, thereby obtaining a high-signal-to-noise-ratio baseband quantum signal. Finally, a RRC filter with a roll-off factor of $0.2$ is adopted to perform matched filtering on the baseband signal. After matched filtering and optimum sampling point extraction, high-fidelity quadrature components are output, providing a reliable data foundation for the subsequent parameter estimation and key reconciliation.

\begin{figure*}[!t]
	\centering
	\begin{subfigure}[b]{0.3\textwidth}
		\centering
		\includegraphics[width=\linewidth]{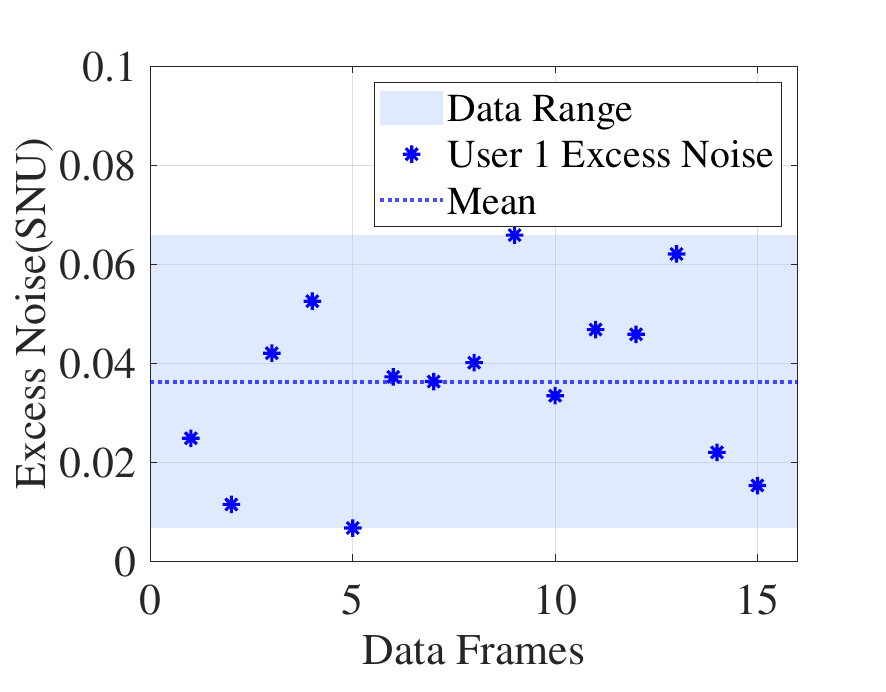}
		\caption{}
		\label{user1}
	\end{subfigure}
	\hfill
	\begin{subfigure}[b]{0.3\textwidth}
		\centering
		\includegraphics[width=\linewidth]{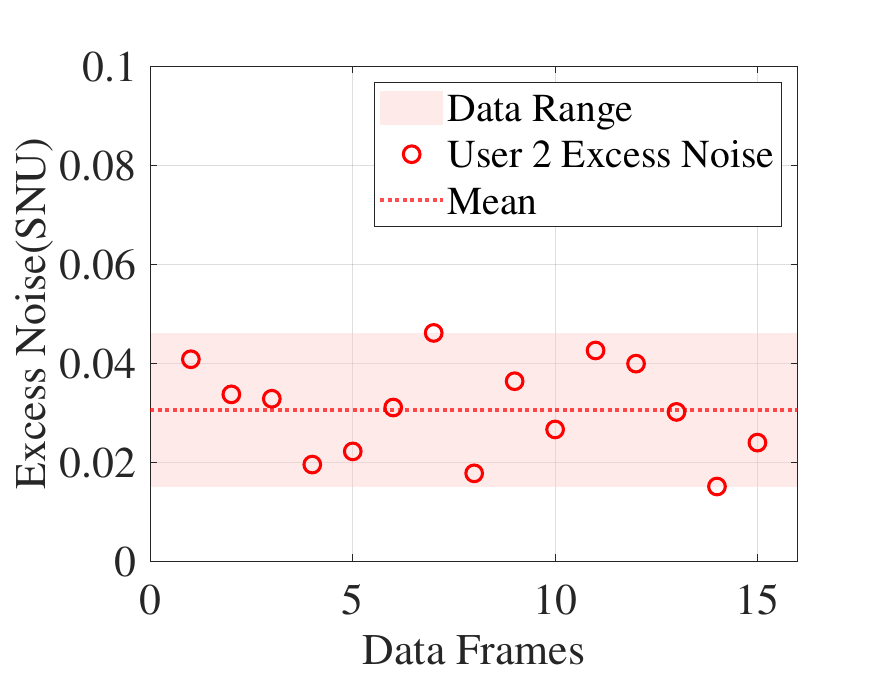}
		\caption{}
		\label{user2}
	\end{subfigure}
	\hfill
	\begin{subfigure}[b]{0.3\textwidth}
		\centering
		\includegraphics[width=\linewidth]{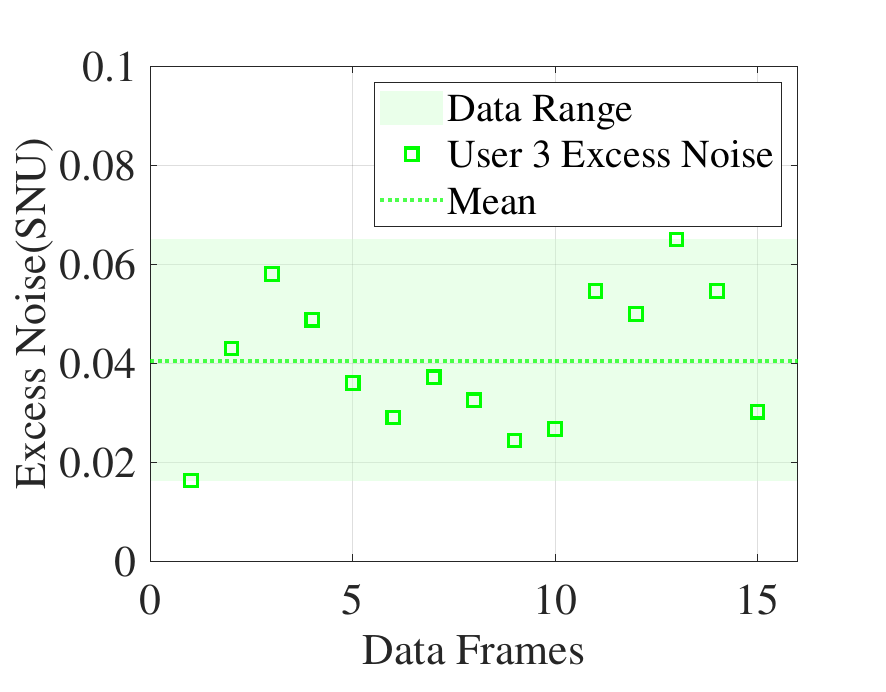}
		\caption{}
		\label{user3}
	\end{subfigure}
	
	\vspace{0.5em} % 添加一些垂直间距
	
	\begin{subfigure}[b]{0.3\textwidth}
		\centering
		\includegraphics[width=\linewidth]{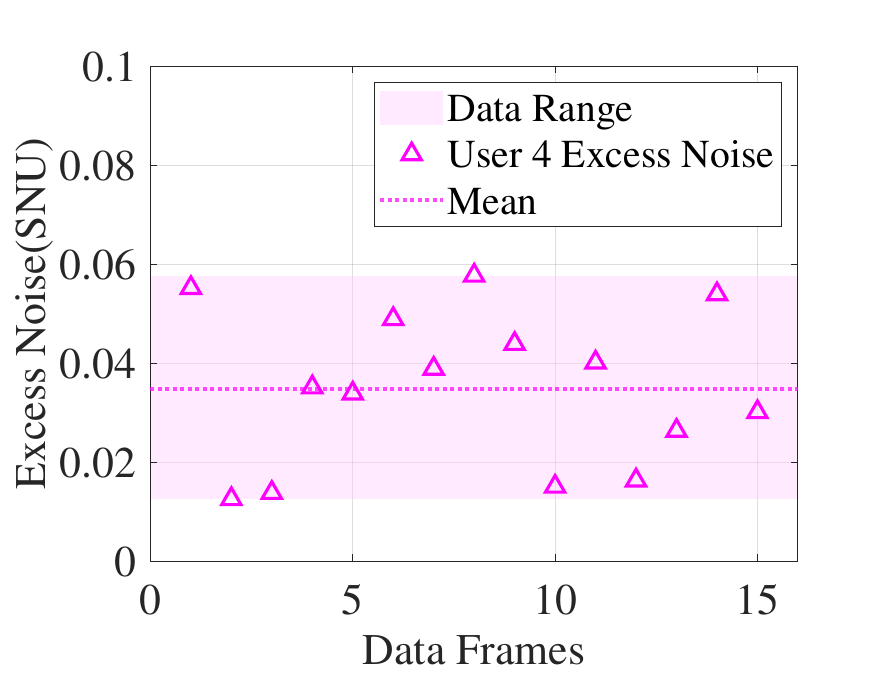}
		\caption{}
		\label{user4}
	\end{subfigure}
	\hfill
	\begin{subfigure}[b]{0.3\textwidth}
		\centering
		\includegraphics[width=\linewidth]{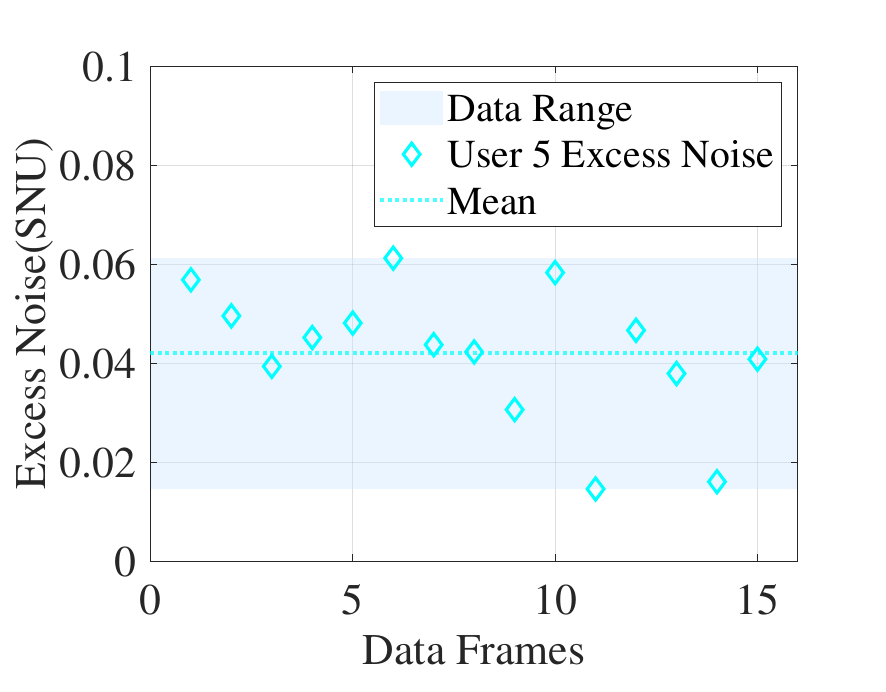}
		\caption{}
		\label{user5}
	\end{subfigure}
	\hfill
	\begin{subfigure}[b]{0.3\textwidth}
		\centering
		\includegraphics[width=\linewidth]{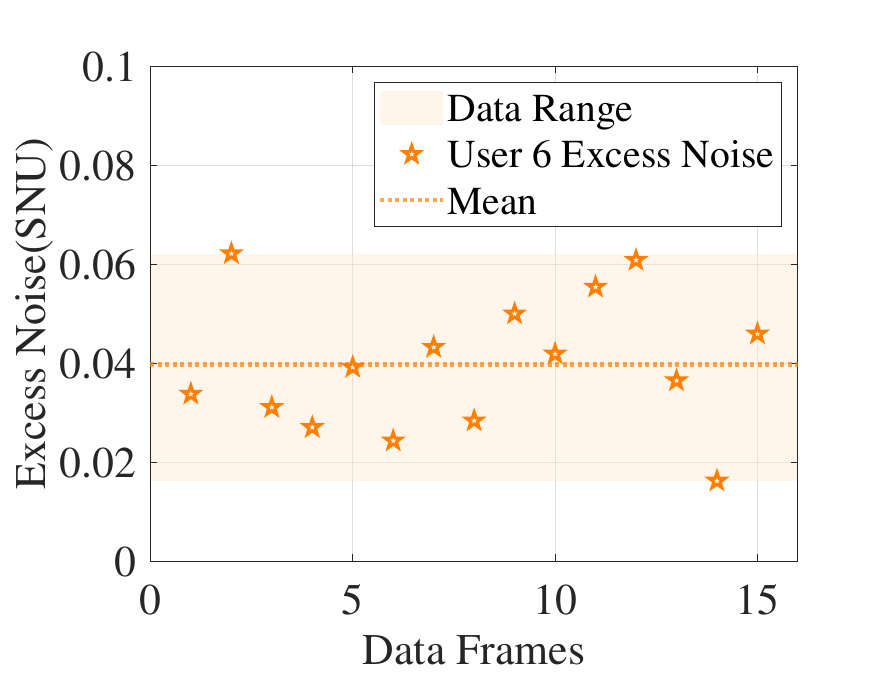}
		\caption{}
		\label{user6}
	\end{subfigure}
	
	\caption{Excess noise data plots for six users. The horizontal dotted line represents the average excess noise for each user. The shaded area indicates the data range. (a) The excess noise data for user 1 is represented by blue asterisks. (b) The excess noise data for user 2 is represented by red dots. (c) The excess noise data for user 3 is represented by green squares. (d) The excessive noise data for user 4 is represented by magenta triangles. (e) The excessive noise data for user 5 is represented by cyan diamonds. (f) The excessive noise data for user 6 is represented by orange pentagrams.}
	\label{fig:excess_noise}
\end{figure*}

\begin{table}[!t]
	\centering
	\caption{System parameters used in the experiment. $A_{ch}$: attenuation; $\varepsilon$: excess noise; $v_{\mathrm{el}}$: electronic noise.}
	\label{tab:system_parameters}
	%\small % 缩小字号
	\setlength{\tabcolsep}{3pt} % 极限压缩列间距以容纳7列
	\begin{tabular}{lcccccc}
		\toprule
		\multicolumn{7}{c}{\textbf{Common Parameters}} \\
		\midrule
		\multicolumn{3}{c}{\textbf{Parameter}} & \multicolumn{4}{c}{\textbf{Value}} \\ % 新增的表头行
		\midrule
		\multicolumn{3}{l}{Symbol rate $f$} & \multicolumn{4}{c}{500 MHz} \\
		\multicolumn{3}{l}{Modulation variance $V_A$} & \multicolumn{4}{c}{4.3 SNU} \\
		\multicolumn{3}{l}{Quantum efficiency $\eta$} & \multicolumn{4}{c}{0.56} \\
		\multicolumn{3}{l}{Reconciliation efficiency $\beta$} & \multicolumn{4}{c}{96\%} \\
		\midrule
		\multicolumn{7}{c}{\textbf{User-Specific Parameters}} \\
		\midrule
		\textbf{Parameter} & \textbf{User 1} & \textbf{User 2} & \textbf{User 3} & \textbf{User 4} & \textbf{User 5} & \textbf{User 6} \\
		\midrule
		$A_{ch}$ (dB) & 15.00 & 14.67 & 15.92 & 16.16 & 16.78 & 16.48 \\
		$\varepsilon$ (SNU) & 0.0361 & 0.0306 & 0.0404 & 0.0348 & 0.0421 & 0.0397 \\
		$v_{\mathrm{el}}$ (SNU) & 0.1612 & 0.1624 & 0.1597 & 0.1627 & 0.1637 & 0.1627 \\
		\bottomrule
	\end{tabular}
\end{table}

\subsection{Experimental results}
In particular, the shared modulation variance in the experiment is chosen as $V_A=4.3$SNU. According to the max-min criterion introduced in Section \ref{subsubsec:Min-SKR}, this value serves as the optimal shared operating point for the considered asymmetric field-deployed access scenario, and is therefore adopted as the common system parameter for all users.

\subsubsection{Parameter evaluation}
Parameter evaluation is performed using 15 data frames per user, with each frame containing $10^7$ symbols. The key system parameters, including the modulation variance, quantum efficiency, and user-specific electronic noise, are summarized in Table~\ref{tab:system_parameters}. Fig.~\ref{fig:excess_noise} illustrates the evaluated excess noise across the six users. In the scatter plots, individual points represent the specific excess noise for each data frame, while the shaded areas denote the data range. The horizontal dotted lines indicate the 15-frame average excess noise ($\varepsilon$) for each user, which exactly corresponds to the $\varepsilon$ values listed in Table~\ref{tab:system_parameters}.

\subsubsection{Experimental SKR analysis}
As detailed in Eq. (\ref{eq:SKR}), the SKR between Alice and a specific user, $\text{Bob}_i$, in the point-to-multipoint CV-QAN can be simplified as:
\begin{equation}
	\mathrm{SKR}_{AB_i} = f(\beta I_{AB_i} - \chi_{B_iE}).
\end{equation}

\begin{figure}[!t]
	\centering
	\includegraphics[width=8.5cm]{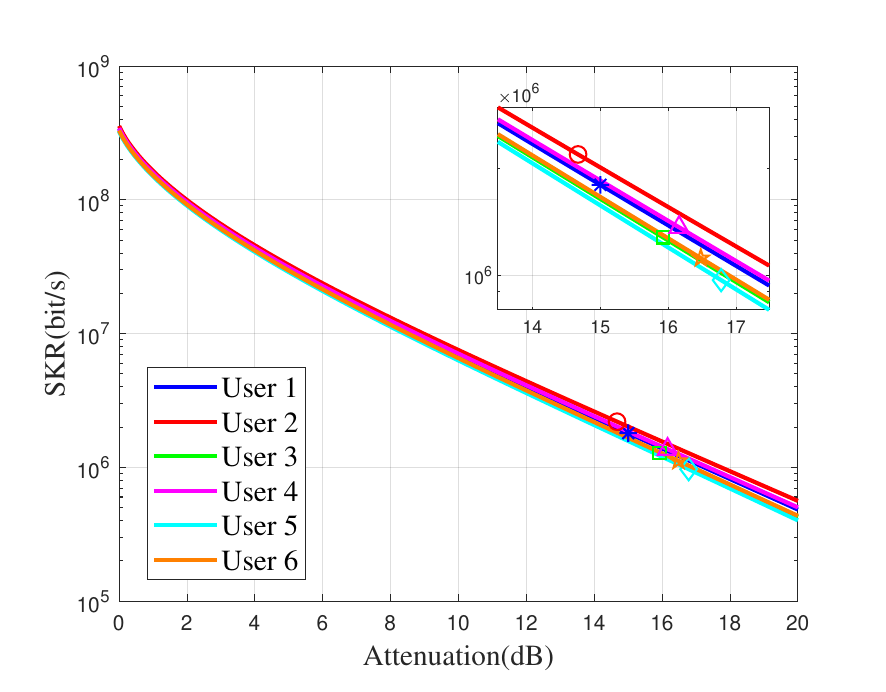}
	\caption{Experimental results of CV-QAN. The blue line, red line, green line, magenta line, cyan line, and orange line represent the SKR curves of user 1, user 2, user 3, user 4, user 5, and user 6, respectively. }
	\label{fig:skr}
\end{figure}

Based on the parameters in Table~\ref{tab:system_parameters}, the theoretical asymptotic SKR curves for the six users are plotted as solid lines in Fig.~\ref{fig:skr}. The inset in the top right provides a magnified view of the experimental results. The experimentally obtained SKRs for users 1 through 6 are 1.80, 2.20, 1.28, 1.37, 0.97, and 1.12 Mbit/s, respectively.

\subsubsection{Composable security analysis}
We further analyze the SKR under the composable-security framework. The corresponding SKR is calculated as follows \cite{wang2025high}:
\begin{equation}
	\mathrm{SKR}^{CS}_{AB_i}=\frac{1-\mathrm{FER}}{N}[n\mathrm{SKR}_{AB_i}-\sqrt{n}\Delta_{aep}+\log_2(2\varepsilon_h^2\varepsilon_{cor})-1].
\end{equation}
where $n$ represents the effective key generation length, and $N$ is the total block length, of which only a fraction $n/N$ is utilized for key generation. The parameter $\mathrm{FER}$ denotes the frame error rate of the error correction process, and $\mathrm{SKR}_{AB_i}$ is the corresponding asymptotic SKR. The security parameters $\varepsilon_h$ and $\varepsilon_{cor}$ correspond to the bounds for privacy amplification and error verification, respectively. To account for finite-size effects, the penalty term $\Delta_{aep}$, derived from the asymptotic equipartition property (AEP) for bounding the smooth-min entropy, is calculated as:
\begin{equation}
	\label{eq:delta_aep}
	\Delta_{\text{aep}} = 4 \log_2(\sqrt{2^d} + 2)\sqrt{-\log_2(1-\sqrt{1-\varepsilon_s^2})}.
\end{equation}
Here, $d$ represents the quantization resolution (in bits) of the ADC at the receiver's side, and $\varepsilon_s$ is the smoothing parameter associated with the estimation of the smooth-min entropy of the quantum state.

\begin{figure}[!t]
	\centering
	\includegraphics[width=8.5cm]{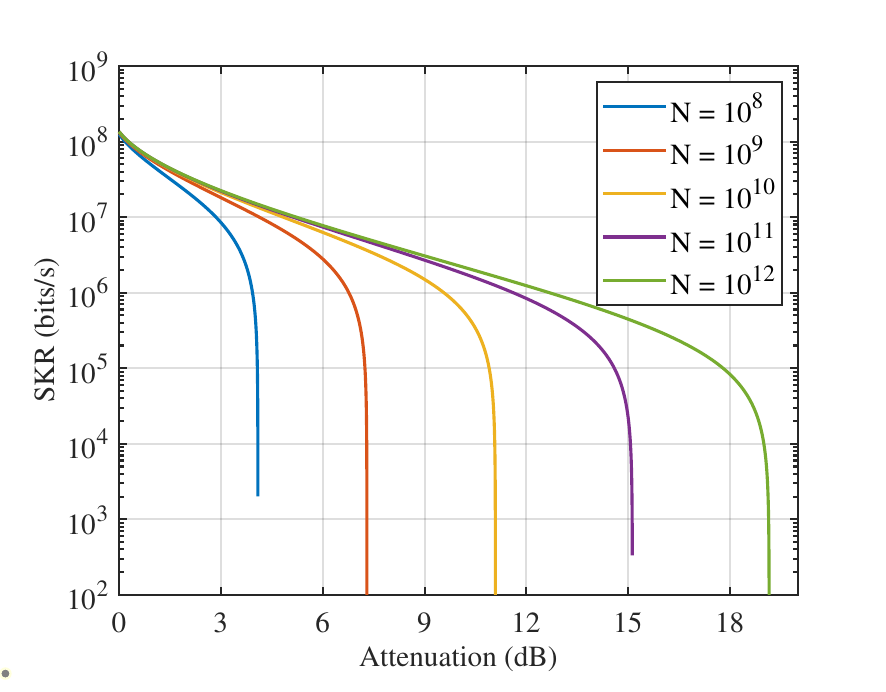}
	\caption{Composable security SKR simulation diagram. Simulation calculations were performed for user 1's composable security SKR curve with block lengths $N = 10^8,10^9,10^{10},10^{11}$ and $10^{12}$.}
	\label{fig:skr_N}
\end{figure}

During the simulation, we set $\varepsilon_h=\varepsilon_{cor}=\varepsilon_s=10^{-10}$, $\mathrm{FER}=0.2$, $d=8$,and $n/N=0.9$. As a representative example, we evaluate the composable-security SKR of User 1 for block lengths $N = 10^8,10^9,10^{10},10^{11}$ and $10^{12}$, as shown in Fig. \ref{fig:skr_N}. 
The simulation is performed to illustrate the finite-size requirement under the theoretical 12 dB splitting loss of the 1-to-16 access architecture. The results suggest that, under this condition, a block length of approximately $N=10^{11}$ is needed to achieve a positive composable key rate.

\section{Discussion}
While research on QKD networks is progressively transitioning toward end-user access, practical QANs have largely remained confined to laboratory conditions . Targeting realistic optical fiber
access environments, we present a field demonstration of a downstream CV-QAN and further address the critical problem of selecting the system’s modulation variance across asymmetric channels

Theoretically, we address the inherent challenges of asymmetric optical fiber links. Because the modulation variance ($V_A$) in a broadcast architecture is a system-wide shared parameter, the varying lengths of access fibers mean the optimal operating point naturally differs per user. Our analytical framework demonstrates that under asymmetric conditions, the optimal $V_A$ varies significantly depending on the specific optimization objective. Consequently, selecting $V_A$ can no longer rely on symmetric models; instead, it necessitates a deliberate trade-off between overarching system efficiency and user fairness.

Experimentally, we deployed the downstream broadcast CV-QAN over existing commercial telecommunication fiber infrastructure, constructing a tree-topology access network tailored for the QTTH scenario. By integrating essential DSP procedures, we achieved field validation of the multi-user access network. The experimental results demonstrate that even under the practical conditions of asymmetric link lengths, each user is still capable of achieving an SKR at the Mbit/s level, indicating that the broadcast CV-QAN shows promising engineering feasibility for deployment in real-world access environments.

From an application perspective, the downstream broadcast CV-QAN exhibits high compatibility with traditional passive optical networks, establishing a practical foundation for extending quantum-secure capabilities to diverse end-user scenarios. However, propelling this technology toward widespread practical application requires addressing critical challenges in large-scale resource allocation, system control, and long-term stability.

Crucially, compared to laboratory settings, the system noise sources in deployed optical fiber access networks are inherently more complex, exhibiting significantly higher overall fluctuation levels. Under such conditions, although this study has successfully verified the feasibility of the multi-user broadcast CV-QAN in a field environment, conducting a rigorous security analysis under the finite-size regime remains a nontrivial task. Consequently, future work will focus on acquiring data with larger block sizes and achieving prolonged stable operation, progressively advancing the system toward rigorous verification under finite-size and composable security frameworks.

In summary, while prior research laid the groundwork for multi-user quantum networks, this work extends multi-user quantum access networks to a field-deployed telecommunication environment.By coupling rigorous asymmetric channel modeling with a comprehensive field demonstration, we provide field evidence for the practical feasibility of the broadcast CV-QAN for QTTH scenarios.
\section{Conclusion}
Targeting the QTTH scenario, we present a field demonstration of a downstream broadcast CV-QAN. To address the asymmetric links prevalent in realistic access environments, we establish a comprehensive multi-user analytical framework, which reveals that optimizing the shared modulation variance ($V_A$) inherently necessitates a deliberate trade-off between overall system efficiency and user fairness.

Experimentally, we constructed this tree-topology network over commercial telecommunication infrastructure. Field verifications confirm that the representative users attain SKRs at the Mbit/s level. This provides field evidence for the practical feasibility of broadcast CV-QANs in access-network scenarios, providing empirical evidence for extending quantum-secure capabilities directly to the access network edge.

%%%%%%%%%%%%% Acknowledgements %%%%%%%%%%%%%
\footnotesize
\section*{Acknowledgements}
This work is supported by National Natural Science Foundation of China (No. 62571316, 61971276), Quantum Science and Technology-National Science and Technology Major Project (No. 2021ZD0300703), Shanghai Municipal Science and Technology Major Project (No. 2019SHZDZX01), Natural Science Foundation of Shanghai (No. 25ZR1402251), and Cultivation Project of Shanghai Research Center for Quantum Sciences (No. LZPY2024).).

%%%%%%%%%%%%%%   Bibliography   %%%%%%%%%%%%%%
\normalsize
\bibliography{ref}

%%%%%%%%%%%%  Supplementary Figures  %%%%%%%%%%%%
% \clearpage

%%%%%%%%%%%% Supplementary Methods %%%%%%%%%%%%
\footnotesize

%%%%%%%%%%%%%%%%   End   %%%%%%%%%%%%%%%%
%\end{multicols}  % Method B for two-column formatting (doesn't play well with line numbers), comment out if using method A
\end{document}